\documentclass{aa}

\usepackage{graphicx}
\usepackage{txfonts}
\usepackage[hidelinks]{hyperref}
\usepackage{amsmath, amsfonts, amssymb, slashed, upgreek, color}
\usepackage{natbib}
\usepackage{ulem}
\makeatletter

\renewcommand*{\aa@pageof}{Page \thepage{} of \pageref*{LastPage}}

\makeatother
\makeatletter
\newcommand*{\rom}[1]{\expandafter\@slowromancap\romannumeral #1@}
\makeatother
\makeatletter
\newcommand*{\srom}[1]{\romannumeral #1}
\makeatother
\begin{document}
    \title{Gravitational waves from dark domain walls}

    \author{{\O{}}. Christiansen \inst{1,3}
    \thanks{oyvinch@fzu.cz}
    \and J. Adamek\inst{2}
    \thanks{julian.adamek@uzh.ch}
    \and F. Hassani\inst{1}
    \thanks{farbod.hassani@astro.uio.no}
    \and D. Mota\inst{1}
    \thanks{d.f.mota@astro.uio.no}
    \fnmsep }

    \institute{Institute of Theoretical Astrophysics, University of Oslo, Sem~S\ae{lands}~vei~13, 0371 Oslo, Norway\\ \and Institut f\"ur Astrophysik, Universit\"at Z\"urich, Winterthurerstrasse~190, 8057 Z\"urich, Switzerland \\ \and {CEICO, Institute of Physics of the Czech Academy of Sciences, Na Slovance 1999/2, 182 00, Prague 8, Czechia} }

    \date{}


    \abstract{ For most of cosmic history, the evolution of our Universe has
    been governed by the physics of a `dark sector', consisting of dark matter and
    dark energy, whose properties are only understood in a schematic way. The influence
    of these constituents is mediated exclusively by the force of gravity,
    meaning that insight into their nature must be gleaned from gravitational
    phenomena. The advent of gravitational-wave astronomy has revolutionised the
    field of black hole astrophysics, and opens a new window of discovery for
    cosmological sources. Relevant examples include topological defects, such as
    domain walls or cosmic strings, which are remnants of a phase transition.
    Here we present the first simulations of cosmic structure formation in which
    the dynamics of the dark sector introduces domain walls as a source of
    stochastic gravitational waves in the late Universe. We study in detail how the
    spectrum of gravitational waves is affected by the properties of the model,
    and extrapolate the results to scales relevant to the recent evidence for a
    stochastic gravitational wave background. Our relativistic implementation of
    the field dynamics paves the way for optimal use of the next generation of gravitational
    experiments to unravel the dark sector. }


    \maketitle
    %

    \section{Introduction}
    The detection of the nanohertz stochastic gravitational wave background by
    the pulsar timing arrays \citep{agazie_nanograv_2023,antoniadis_second_2023,reardon_search_2023,xu_searching_2023}
    opens a new window for studying gravitational systems at cosmological scales.
    The signal observed by the NANOGrav array \citep{agazie_nanograv_2023} consists
    of correlated timing residuals of 68 millisecond pulsars that have now been measured
    over a 15-year time period. The amplitude of the timing residuals scales as
    $h_{c}^{2} f^{-3}$, where $f$ is the frequency and $h_{c}$ is the characteristic
    strain of the stochastic gravitational wave background \citep{maiorano_principles_2021},
    i.e.\ the characteristic fractional change in the length of a hypothetical
    detector arm. An astrophysical background from supermassive black hole inspirals
    is predicted \citep{phinney_practical_2001} to yield $h_{c} \sim f^{-2/3}$
    and hence a spectrum of timing residuals that tilts as $f^{-13/3}$. While supermassive
    black holes remain a viable explanation of the observed signal, a tilt
    between $f^{-2.6}$ and $f^{-3.8}$ is preferred by the data, which is in mild
    tension with the prediction. This has prompted a large number of
    investigations into alternative explanations, exploring possible phenomena in
    the dark sector or in the early Universe \citep{agazie_nanograv_2023,babichev_nanograv_2023}.

    One hypothetical source of gravitational waves are topological defects that
    are generically produced at certain types of phase transitions. Depending on
    the nature of the spontaneously broken symmetry, the defects can take the
    form of cosmic strings or domain walls \citep{kibble_topology_1976,saikawa_review_2017,vilenkin_cosmic_1985,vachaspati_kinks_2007}.
    In particular, if the vacuum manifold splits into two (or more) isolated points,
    domain walls separate the regions that have settled into different vacua
    during the phase transition. The case where the phase transition happens in the
    early Universe at a high energy scale has been studied thoroughly in the recent
    literature \citep{ramazanov_beyond_2022,blasi_axionic_2023,li_probing_2023,babichev_nanograv_2023,kitajima_gravitational_2023,gouttenoire_domain_2023,kitajima_stability_2023},
    owing to the dimensional argument that frequencies in the nanohertz range today
    correspond to the size of the causal horizon (the scale that sets the
    typical size of different vacuum domains) when the Universe had a temperature
    of several trillion degrees Kelvin.

    In this work instead, we consider domain walls appearing in the late-time
    Universe due to phase transitions in the dark energy field. This scenario has
    not yet been studied in the context of stochastic gravitational waves. We
    consider the (a)symmetron model \citep{perivolaropoulos_gravitational_2022},
    a generalisation of the symmetron \citep{hinterbichler_symmetron_2010,hinterbichler_symmetron_2011},
    a theory of gravity with an extra scalar degree of freedom. Such extra
    degrees of freedom are a general prediction of physics beyond the Standard
    Model \citep{clifton_modified_2012}. The symmetron can be regarded as a
    generic scalar effective field theory (EFT), with a potential parameterized as
    a quartic-order Taylor polynomial around the field origin. The quartic self-coupling
    term is chosen to be positive so that the potential is bounded from below, while
    the quadratic term is chosen to be negative in order to have a Higgs-like phase
    transition. The $\mathbb{Z}_{2}$ symmetry protects against large quantum
    corrections \citep{hinterbichler_symmetron_2010}. The symmetron has a
    universal coupling to the matter sector through a conformal factor, which adds
    a coupling term to the effective mass of the scalar field. The model can
    thus be seen as a typical EFT limit for $\mathbb{Z}_{2}$ symmetric Horndeski
    theories in the case where the field $\phi/M_{\rm{pl}}$ is small so that the
    Taylor expansion can be truncated. In addition, we have added an odd term to
    the potential that explicitly breaks the $\mathbb{Z}_{2}$ symmetry, so that we
    can study the case where the vacuum degeneracy is lifted and one of the minima
    is favoured.

    The universal coupling to the matter sector mediates a fifth force that affects
    cosmological structure formation and is tightly constrained by local gravity
    experiments \citep{will_confrontation_2014,bertotti_test_2003,esposito-farese_tests_2004,burrage_accurate_2023,llinares_detecting_2019}.
    When the local matter density falls below some critical threshold $\rho_{*}$,
    it triggers the $\mathbb{Z}_{2}$ phase transition, and the field settles into
    either of two minima. A crucial difference from the early Universe is that
    the late-time Universe has very large fluctuations in the local density, so that
    the phase transition happens in different places at different times. The
    resulting domains are separated by domain walls that are moulded by the
    sheets and filaments in the cosmic large-scale structure. This way, a large
    number of domain walls can be formed in a Hubble volume. Both their energy density
    and Compton scale are free parameters, but the former is often chosen well
    below the critical density such that the expansion rate of the Universe is
    not affected significantly. However, they can still produce a considerable
    stochastic gravitational wave signal if their time evolution is violent
    enough. In the case where the potential is not symmetric, the asymmetry
    tends to destabilise the domains that occupy the higher minimum. Spontaneous
    reconfiguration of domains then leads to the collision and annihilation of domain
    walls, adding a further source of gravitational waves that is qualitatively
    different from stable domain walls.

    In addition to being considered as a model for dark matter \citep{burrage_symmetron_2019,kading_lensing_2023},
    the symmetron can also be motivated by the recent results of the Dark Energy
    Spectroscopic Instrument (DESI), which hint at a dynamical source for late-time
    cosmic acceleration \citep{desi_collaboration_desi_2024}. In recent work \citep{christiansen_environmental_2024}
    it was shown that a consistent evolution of the cosmic acceleration can be
    found using the symmetron together with a degravitation mechanism. Alternatively,
    one can still imagine obtaining a consistent accelerated expansion from the
    combination of the slow-rolling scalar with decreasing potential energy (and
    equation of state parameter $w \sim -1$) and the increasingly dominant
    domain walls ($w \sim -2/3$). The symmetron has also been linked to the
    detection of an anomalous Integrated Sachs-Wolfe (ISW) signal \citep{christiansen_asimulation_2024, kovacs_more_2019},
    and to claims of large structure being in tension with the cosmological principle
    \citep{secrest_test_2021,secrest_challenge_2022,peebles_flat_2023,lopez_big_2024,nadathur_seeing_2013}.

    \begin{figure*}
        \centering
        \includegraphics[width=\linewidth]{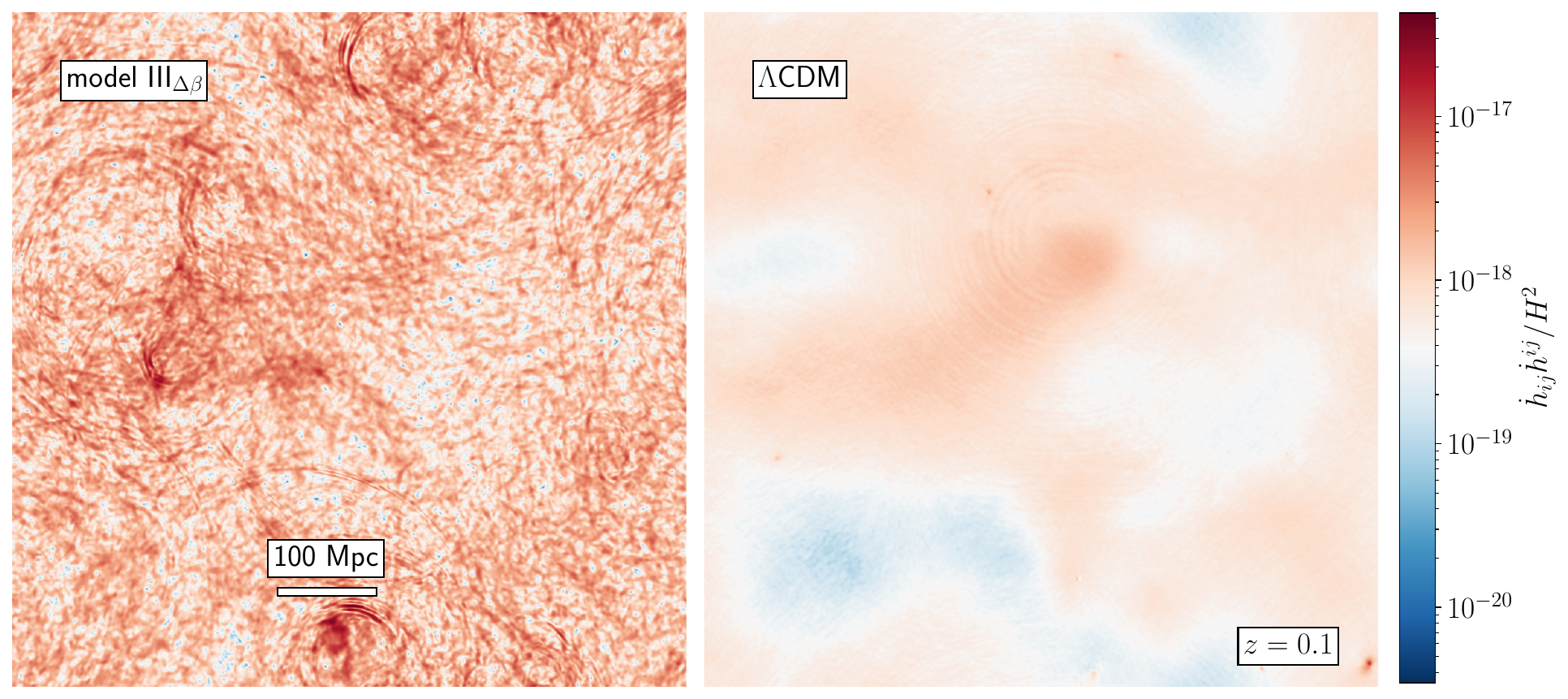}
        \caption{A slice through the simulation volume at redshift $z=0.1$, showing
        the tensor wave intensity $\sum_{i,j}\dot{h}^{2}_{ij}/H^{2}$, where
        ${H}$ is the Hubble factor. Model \rom{3}$_{\Delta \beta}$, shown on the
        left, has the parameters ($L_{C},z_{*},\bar\beta,\Delta\beta/\bar\beta)=(
        1480$
        kpc$, 0.1,8,10\%$). A standard cosmology without the scalar field is shown
        on the right. (The Supplementary Material contains animations that show the
        evolution of the fields and the formation of cosmic structure.) }
        \label{fig:nicefigure}
    \end{figure*}

    \section{Gravitational wave spectra}
    We write the space-time line element of the expanding Friedmann universe with
    linearised gravitational waves as
    \begin{equation}
        \label{eq:metric}ds^{2} = a^{2}(\tau) \left[-c^{2} d\tau^{2} + \left(\delta
        _{ij}+ h_{ij}\right) dx^{i} dx^{j}\right]\,,
    \end{equation}
    where $a$ denotes the scale factor, $c$ is the speed of light, $d\tau = a^{-1}
    dt$ is a conformal time element, $x^{i}$ are comoving Cartesian coordinates
    on the spatial hypersurface, and $h_{ij}$ is the transverse and traceless tensor
    perturbation containing the gravitational waves. A sum is implied over
    repeated indices. Our simulations also track scalar and vector perturbations
    of the metric \citep{adamek_general_2016}, but these are relevant mainly for
    matter dynamics and shall be omitted in the following discussion for brevity.
    In Fourier space, the tensor perturbations obey a damped oscillator equation
    that is driven by the spin-2 projection of the stress-energy tensor \citep{adamek_gevolution_2016},
    \begin{equation}
        \label{eq:waveequation}\ddot{\tilde{h}}_{ij}+ 2 \mathcal{H}\dot{\tilde{h}}
        _{ij}+ c^{2} k^{2} \tilde{h}_{ij} = \frac{16 \pi G}{c^{2}}\left(P_{i}^{l}
        P_{j}^{m} - \frac{1}{2}P_{ij}P^{lm}\right) \tilde{T}_{lm}\,,
    \end{equation}
    where a dot denotes a derivative with respect to $\tau$,
    $\mathcal{H}= d\ln a/d\tau$ is the conformal Hubble rate, $G$ is Newton's gravitational
    constant, $k$ is the comoving Fourier wavenumber, and $P_{ij}= \delta_{ij}- k
    ^{-2}k_{i} k_{j}$ is a transverse projector. A tilde indicates spatial
    Fourier transform, and $\tilde{T}_{lm}$ is the spatial part of the stress-energy
    tensor in Fourier space.

    In our numerical simulations, we compute the stress-energy tensor of the
    scalar field, and hence of the domain walls, non-perturbatively by solving
    the full equations of motion discretised on a three-dimensional grid. The stress-energy
    is taken to Fourier space, using fast Fourier transforms, to source the wave
    equation \eqref{eq:waveequation} that we integrate separately for each wave
    vector using a temporal leapfrog integrator. We evolve only the Fourier components
    for which our time stepping guarantees a stable integration, which requires
    $k \lesssim (c \Delta\tau)^{-1}$ for an integration step $\Delta\tau$. At any
    point in time, we can estimate the power spectrum $P_{\dot{h}}(k)$ of
    $\dot{h}_{ij}$,
    \begin{equation}
        \sum_{i,j}\left\langle\dot{\tilde{h}}_{ij}(\boldsymbol{k}, \tau) \dot{\tilde{h}}
        _{ij}(\boldsymbol{k}', \tau)\right\rangle = \delta^{(3)}_{\mathrm{D}}(\boldsymbol
        {k}+ \boldsymbol{k}') P_{\dot{h}}(k)\,,
    \end{equation}
    where $\delta^{(3)}_{\mathrm{D}}$ is the three-dimensional Dirac delta distribution.
    Assuming that the gravitational waves propagate freely at the time when we
    perform the measurement, this spatial two-point correlation is directly related
    to the temporal one via the retarded time. In this case one finds that the
    spectral density in the frequency domain, $S_{h}(f)$, is simply given by
    $2 \pi c^{3} S_{h}(f) = P_{\dot{h}}(k = 2 \pi f / c)$.

    It is common practice to express the spectrum in terms of the energy density
    of gravitational waves \citep{allen_detecting_1999},
    \begin{multline}
        \rho_{\mathrm{gw}}= \int\limits_{0}^{\infty} \frac{d\rho_{\mathrm{gw}}}{df}
        df = \frac{c^{2}}{32 \pi G}\sum_{i,j}\left\langle \dot{h}_{ij}(\boldsymbol
        {x},\tau) \dot{h}_{ij}(\boldsymbol{x},\tau)\right\rangle\\ = \frac{\pi c^{2}}{2
        G}\int\limits_{0}^{\infty} f^{2} S_{h}(f) df\,.
    \end{multline}
    One may furthermore define the dimensionless density per logarithmic
    frequency interval,
    \begin{equation}
        \Omega_{\mathrm{gw}}(f) = \rho_{\mathrm{crit}}^{-1}\frac{d\rho_{\mathrm{gw}}}{d\ln
        f}= \frac{4 \pi^{2}}{3 H^{2}_{0}}f^{3} S_{h}(f) = \frac{2 \pi^{2}}{3 H^{2}_{0}}
        f^{2} h_{c}^{2}(f)\,,
    \end{equation}
    where $H_{0}$ is the present-day Hubble parameter and the critical density of
    the Universe is given by $\rho_{\mathrm{crit}}= 3 c^{2} H_{0}^{2} / (8 \pi G)$.

    \section{Numerical simulations} \label{S:numerical_simulations}

    The asymmetron model depends on four parameters \citep{christiansen_asevolution_2023}:
    the Compton wavelength $L_{C}$, indicating the correlation length of the scalar
    field in vacuum; the cosmological redshift of the phase transition $z_{*}$, indicating
    the time when the background matter density is equal to the critical density
    $\rho_{*}$ of the phase transition; the average strength of the fifth force
    relative to gravity, $\bar\beta = (\beta_{+} + \beta_{-})/2$; and the
    respective difference $\Delta\beta = \beta_{+} - \beta_{-}$ in the two
    minima. The asymmetron reduces to the symmetron for the choice $\Delta\beta=0$,
    in which case the scalar potential is symmetric.

    We run two suites of simulations\footnote{The code \citep{christiansen_asimulation_2024,christiansen_asevolution_2023,adamek_gevolution_2016,daverio_latfield2_2016,llinares_releasing_2013,llinares_cosmological_2014}
    is publicly available and can be downloaded at \url{https://github.com/oyvach/AsGRD}.}
    varying parameters about their respective values in a reference model.
    Within the first suite, three high-resolution simulations ($1280^{3}$ grid
    points, $742$ Mpc box size) are labelled \rom{1}-\rom{3}, and two lower-resolution
    simulations ($512^{3}$ grid points, $742$ Mpc box size) are labelled \rom{4}
    and \rom{5}. The five simulations within the second suite roughly match the higher
    resolution but use a smaller volume ($256^{3}$ grid points, $148$ Mpc box size),
    and are labelled \srom{1}-\srom{5}. Among the larger volume simulations, the
    reference model \rom{1} has parameters ($L_{C},z_{*},\bar\beta,\Delta\beta)=(
    1480\,\mathrm{kpc}, 0.1,8,0$). Models \rom{2}-\rom{5} vary one parameter at a
    time, using $L_{C}=1111\,\mathrm{kpc}$, $\Delta\beta/\bar\beta = 10 \%$,
    $z_{*}=0$ and $\bar \beta=12$, respectively. For quick reference, we use a subscript
    to indicate the parameter that was varied with respect to the reference model.
    In the smaller volume simulations, our reference model \srom{1}\ has ($L_{C},
    z_{*},\bar\beta,\Delta\beta)=(743\,\mathrm{kpc}, -0.2,90,0$), and models \srom{2}-\srom{5}\ use
    $L_{C}=801\,\mathrm{kpc}$, $\Delta\beta/\bar\beta=30\%$, $z_{*} = -0.13$ and
    $\bar\beta=120$, respectively. The choices of Compton wavelength are close to
    the resolution limit of these simulations{, and we comment more on numerical resolution and convergence in appendix \ref{A:convergence}}.
    Smaller box sizes, allowing correspondingly
    smaller Compton wavelengths, become increasingly expensive to simulate due
    to the shorter dynamical timescales involved. The values of $\beta$ chosen are
    the largest that satisfy the code's linearisation scheme \citep{christiansen_asevolution_2023}.
    The small choices of $z_{*}\lesssim 0$ are largely unconstrained \citep{burrage_accurate_2023}.

    The left panel of figure~\ref{fig:nicefigure} shows the intensity of
    gravitational waves emanating from collapsing domain walls for the model
    \rom{3}$_{\Delta\beta}$ that has an asymmetry of $10\%$ in the scalar field
    potential. In contrast, the right panel shows the gravitational wave pattern
    for a standard $\Lambda$-cold-dark-matter ($\Lambda$CDM) cosmological model without
    the scalar field. In this case, the main source of long-wavelength gravitational
    waves is the evolving large-scale structure.

    \begin{figure*}
        \centering
        \includegraphics[width=\linewidth]{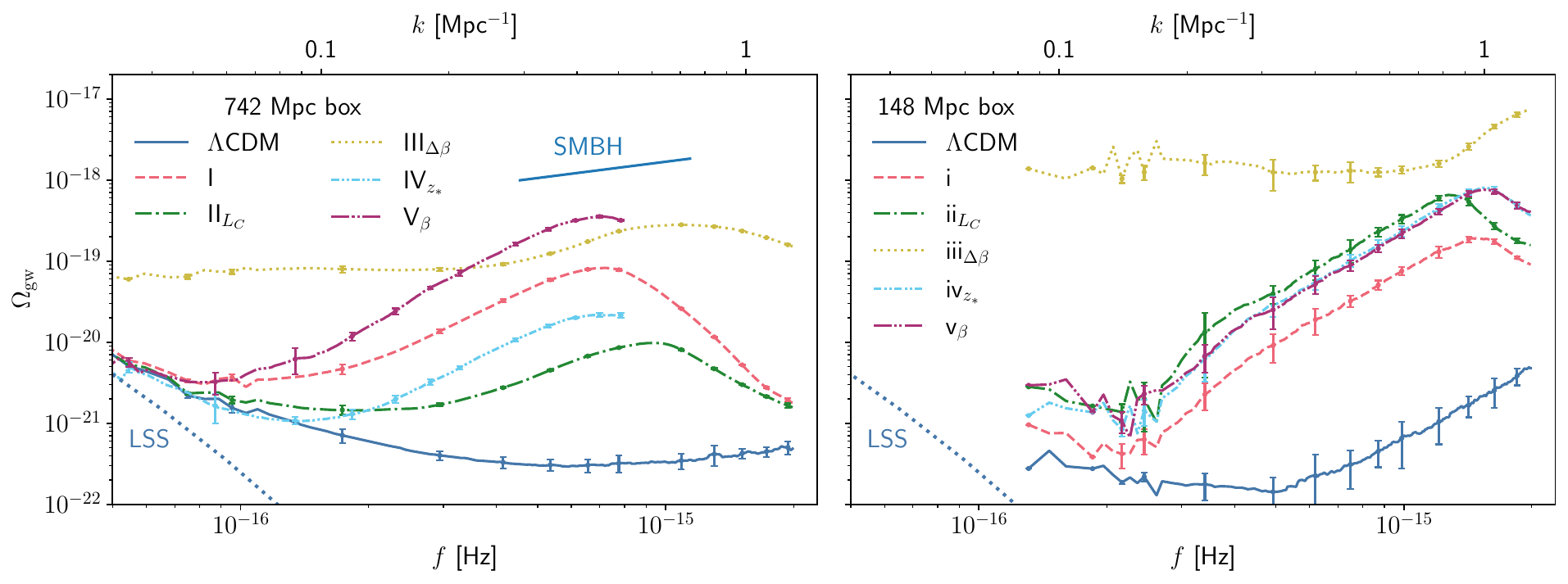}
        \caption{Power spectra of the time derivative of the tensor perturbation,
        $\dot{h}_{ij}$, plotted as dimensionless energy density per logarithmic
        frequency interval. The conversion from Fourier wavenumber $k$ to frequency
        $f$ assumes the standard dispersion relation $k = 2 \pi f / c$. Results
        from the simulations in the larger volume, $(742\,\mathrm{Mpc})^{3}$,
        are shown in the left panel and the ones for the smaller volume, $(148\,\mathrm{Mpc}
        )^{3}$, are shown in the right panel. The error bars indicate the 95\%
        confidence intervals estimated from statistical fluctuations measured
        for each realisation within a narrow frequency band. The dotted line labelled
        ``LSS'' shows the gravitational wave contribution from the evolution of large-scale
        structure as predicted by second-order perturbation theory. The short
        blue line labelled ``SMBH'', inserted as a visual guide only, indicates
        the expected slope for an astrophysical signal produced by supermassive
        black hole inspirals. This signal would however appear at very different
        frequencies and amplitudes. }
        \label{fig:hijprime_comparison_pk}
    \end{figure*}

    Figure~\ref{fig:hijprime_comparison_pk} shows the present-day ($z=0$) power
    spectra of gravitational waves, presented in units of dimensionless energy density
    per logarithmic frequency interval. There is a clear generation of power at
    the scale corresponding to the Compton wavelength of the theory with a power-law
    tail extending to lower frequencies. At extremely low frequency, below about
    $10^{-16}\,\mathrm{Hz}$, cosmic large-scale structure (LSS) becomes the dominant
    source of gravitational waves in most cases. The corresponding amplitude can
    be predicted from second-order perturbation theory and is indicated as
    dotted line in the figure, showing good agreement with our simulations in the
    regime where perturbation theory is valid. The somewhat larger amplitude we
    find for $\Lambda$CDM is due to higher-order nonlinearities and, at the
    highest wavenumbers, discretisation noise in our numerical scheme.


    \begin{figure*}
        \centering
        \includegraphics[width=\linewidth]{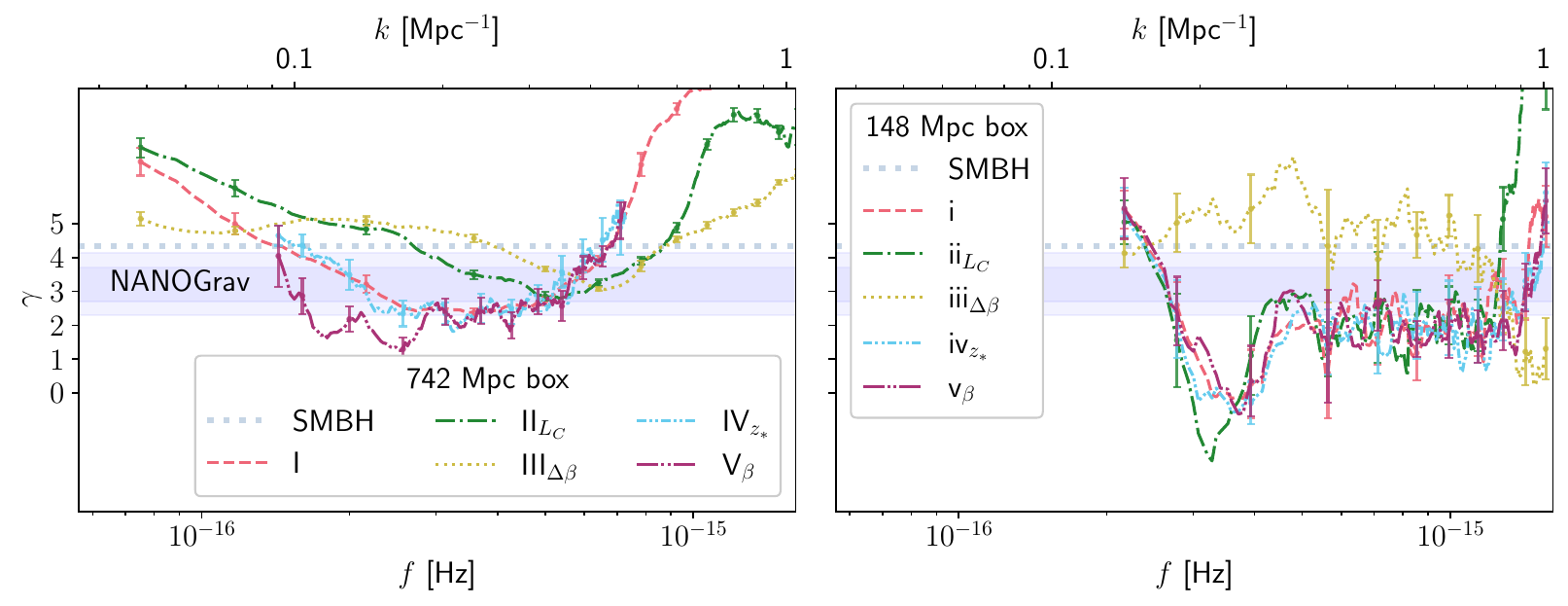}
        \caption{The spectral index $\gamma$ inferred from the power spectra
        shown in figure~\ref{fig:hijprime_comparison_pk}, with error bars indicating
        95\% confidence intervals. The blue bands show the 68\% and 95\% confidence
        intervals reported for the NANOGrav observations that were however taken
        at much higher frequencies. The dotted horizontal line labelled SMBH
        indicates the expected spectral index for supermassive black hole
        inspirals, which would also appear at much higher frequencies. }
        \label{fig:spectral_index}
    \end{figure*}

    The following qualitative trends can be seen in our suite of simulations: a
    smaller Compton wavelength $L_{C}$ reduces the amplitude and moves the peak
    power to a correspondingly smaller scale; a larger average fifth force $\bar\beta$
    increases the amplitude; and a later time of symmetry breaking (lower $z_{*}$)
    reduces the amplitude and has otherwise a small effect. Finally, adding
    asymmetry to the potential by choosing $\Delta\beta>0$, we find more unstable
    domains, significantly increasing the contribution of collapsing domain
    walls to the gravitational wave signal. We find that this component, which
    consists of a noise of gravitational wave bursts, has a nearly flat spectrum
    at low frequencies.

    The NANOGrav collaboration has presented confidence intervals for the spectral
    tilt by fitting a power law $\sim A f^{-\gamma}$ to the timing residuals \citep{agazie_nanograv_2023}.
    We measure the spectral tilt $\gamma$ from our simulations, using the fact
    that the timing residuals scale as $\propto \Omega_{\mathrm{gw}}f^{-5}$. Our
    results are shown in figure~\ref{fig:spectral_index}. We note that all
    simulations display a plateauing spectral tilt that persists up to the
    Compton scale. Furthermore, there is no strong indication for a dependence
    on $\bar \beta$ and $z_{*}$. Stable domain walls seem to generate slightly bluer
    spectra (lower $\gamma$) compared to the NANOGrav measurement, whereas
    collapsing and annihilating domain walls (models with $\Delta\beta >0$) push
    the tilt all the way to the red side of the NANOGrav posterior.

    \section{Discussion}

    To connect our results with the NANOGrav measurements, we need to
    extrapolate eight orders of magnitude towards smaller scales and about ten
    orders of magnitude in energy density $\Omega_{\mathrm{gw}}$. Such a huge extrapolation
    in parameter space comes with several caveats that we address partially, and
    they should be investigated further in the future. To get some analytic insight,
    it is useful to consider that, when viewed on scales larger than the thickness
    of the domain wall, its stress-energy is simply proportional to the surface
    tension
    \begin{equation}
        \sigma \simeq \int_{\phi_-}^{\phi_+}\sqrt{2 V_{\rm{eff}}(\phi) -2V_{\rm{eff}}(\phi_{\pm})}
        \,,
    \end{equation}
    where one integrates between the two asymptotic vacuum values $\phi_{\pm}$
    on the two sides of the wall. At vanishing matter density we have $\sigma=12
    \sqrt{2}\, \Omega_{\mathrm{m}}^{2} H_{0}^{4} \bar{\beta}^{2}(1+z_{*})^{6}L_{C}
    ^{3}$. This sets the strength of the gravitational wave source and we expect
    that $\Omega_{\mathrm{gw}}\propto \sigma^{2}$. Our simulations indeed confirm
    this scaling with $\bar{\beta}$, while we find a slightly stronger-than-expected
    dependence $\Omega_{\mathrm{gw}}\sim (1+z_{*})^{14}$. The dependence on $L_{C}$
    seems much more complicated than the naive consideration of $\sigma$ suggests.
    This may be due to the fact that domain walls tend to be aligned with sheets
    and filaments of dark matter, affecting the effective potential and hence the
    surface tension in a non-trivial way. We point out that the Compton scale
    $L_{C}$ sets the typical length scale over which the field can respond to
    local variations in the matter density.

    Given that $L_{C}$ would need to be reduced by some eight orders of
    magnitude from our reference simulations to reach NANOGrav frequencies, it is
    difficult to predict the amplitude robustly. The trends seen in our
    simulations suggest that the amplitude would need to be boosted
    significantly to reach NANOGrav levels, which can be achieved by increasing $\bar
    {\beta}$ or $z_{*}$, or both, presumably by many orders of magnitude. This
    would yield a model with a very strong fifth force that is however also
    strongly screened. Allowing for asymmetry in the potential ($\Delta\beta > 0$)
    is an interesting scenario to explore, as we have seen that the
    gravitational wave amplitude from collapsing domain walls can easily
    dominate the signal.

    In conclusion, we have shown that a late-time cosmological source may be a viable
    candidate for providing the observed spectral tilt of the NANOGrav experiment,
    although matching the signal amplitude remains an open question that
    requires a significant extrapolation of our results. Nevertheless, this work
    presents for the first time a full simulation of an ultra-light scalar field
    in a late-time cosmological $N$-body code, including a consistent treatment of
    gravity and sourcing of gravitational waves. Our code provides a new
    framework and opens the door to studies of the late-time production of
    gravitational waves from cosmological sources, such as topological defects,
    late-time phase transitions, and their interaction with structure formation.
    This tool is set to play an important role at the interface between
    cosmological surveys and gravitational-wave astronomy, both of which are needed
    to probe the dark sector and advance our understanding of the nature of gravity.

    \subsection*{Acknowledgements}
    \O{}.C.\  thanks Constantinos Skordis and the Central European Institute of
    Cosmology (CEICO) for hosting him during the work on the project. J.A.\ acknowledges
    support from the Swiss National Science Foundation.
    This work was co-funded by the European Union and supported by the Czech Ministry of
    Education, Youth and Sports (Project No. FORTE – CZ.02.01.01/00/22\_008/0004632).
    We thank the Research Council
    of Norway for their support. The simulations were performed on resources
    provided by UNINETT Sigma2 -- the National Infrastructure for High Performance
    Computing and Data Storage in Norway. This work is supported by a grant from
    the Swiss National Supercomputing Centre (CSCS) under project ID s1051.

    \subsection*{Author contributions}

    {\O}.C.\ implemented the model in the code, ran the simulations and did most
    of the analysis. J.A.\ proposed and contributed to the code to obtain the gravitational
    wave signal, and played a leading role in the theoretical interpretation of
    the results. All authors discussed the results and contributed to the writing.

    \subsection*{Materials \& Correspondence}

    Correspondence and requests for materials should be directed to {\O}yvind Christiansen
    (\href{mailto:oyvind.christiansen@astro.uio.no}{oyvind.christiansen@astro.uio.no}).

    \subsection*{Data availability statement}
    The data used for producing the figures are available at the link \url{https://zenodo.org/records/10495599}.

    \subsection*{Code availability statement}
    The code `AsGRD' used to produce the results is available from the maintained
    repository \url{https://github.com/oyvach/AsGRD}. The version of the code
    used in this paper, along with the post-processing tools, may be found at
    the link \url{https://zenodo.org/records/10495599}.


    \appendix

    \section{Numerical convergence}\label{A:convergence}

\begin{figure}
        \centering
        \includegraphics[width=\linewidth]{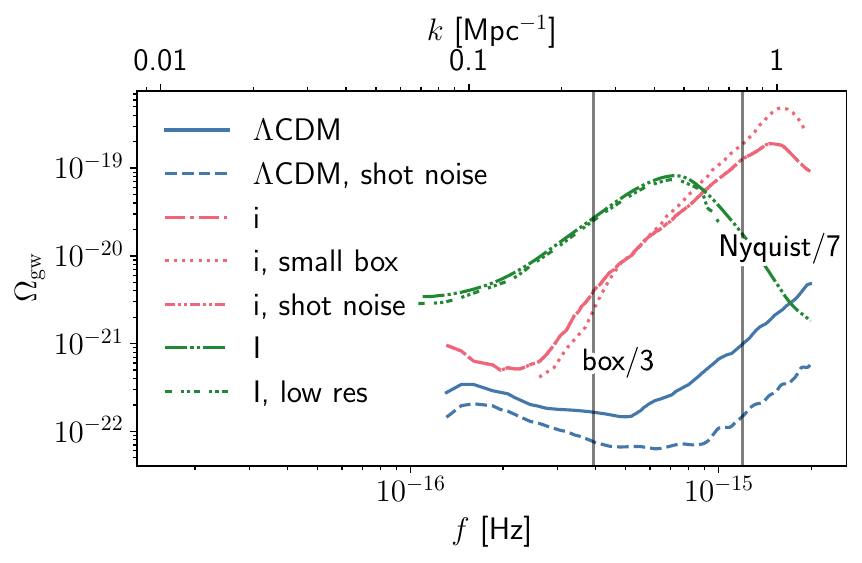}
        \caption{Numerical tests for the power spectra of gravitational
        waves. The red graphs are
        for the parameter choice, \srom{1}, presented in section \ref{S:numerical_simulations},  where we used a simulation volume of (148 Mpc$)^{3}$. The $\Lambda$CDM graphs in blue are using the same box volume. The
        graphs named `shot noise' use 8 times as many particles. The graph
        `small box' keeps the grid and particle number constant, but is using an
        eighth of the simulation volume, $(74$ Mpc$)^{3}$. The green graphs are for the parameter
        choice \rom{1}, and use a simulation volume (148 Mpc$)^3$, but with different grid resolutions of $1280^3$ and $512^3$ points, respectively. The vertical lines indicate the Nyquist scale of the (148 Mpc$)^{3}$  boxes, and the box scale of the  $(74$ Mpc$)^{3}$ box.
        } 
        \label{fig:convergence}
    \end{figure}

    {In order to assess the numerical convergence of our results, we have run 4 additional
    simulations that we present in figure \ref{fig:convergence}}
    {We checked that our results for
    the scalar field models are numerically converged by simulating the fiducial
    model \rom{1} with a lower resolution of only $512^3$ grids.
    This corresponds to the resolutions used for simulations \rom{4}-\rom{5}.
    Figure \ref{fig:convergence} shows perfect agreement between these two cases for
    the relevant range of the spectrum.
    Considering whether we are fully sampling the phase space,} we consider the 
    simulation \srom{1}\, in an eighth of the box volume, keeping the number of
    grid points and particles fixed. We found good agreement for the overlapping
    resolution range, with a loss of power on large scales and a gain of power
    on small scales (above $10^{-15}$\,Hz) in the smaller box, as expected.
    {One should be aware that this smaller simulation is too small to form stable domains
    and has the entire volume collapse into the same domain, and is therefore
    qualitatively different. This may explain the slightly greater slope at large frequencies.} Note additionally
    that the two suites of simulations{, \rom{1}-\rom{5} and \srom{1}-\srom{5}}, shown in
    figure~\ref{fig:hijprime_comparison_pk},
    cover different parameter choices and are therefore expected to be different.
    We tested the robustness of the obtained spectra with respect to the mass discretisation
    by running both the $\Lambda$CDM and the fiducial model \srom{1}\
with $8$ times
    more particles. As a result we saw that the high-frequency rise in the
    $\Lambda$CDM signal from about $f\sim 5\times 10^{-16}$\,Hz is due to discretisation
    noise. For the scalar spectra the noise from the mass discretisation is
    subdominant by several orders of magnitude.

    As we are simulating the sourcing of gravitational waves from domain walls, we comment now in particular
    on how we are resolving them. In a homogeneous background matter density, $\rho$, the static domain wall profile can be solved
    analytically, as is shown e.g.\ in \citet{llinares_domain_2014}, giving
    \begin{align}
        \phi(x)/v_0 &= \sqrt{1-\rho/\rho_*} \tanh \left(
            \frac{a x}{2 L_C}\sqrt{1-\rho/\rho_*}\right)\,,
    \end{align} 
    where $v\equiv \mu/\sqrt{\lambda}$ is the vacuum expectation value of the field $\phi$ and $\rho_*\equiv \mu^2 M^2$ is the critical
    density of the phase transition.
    The width of the profile in conformal coordinates therefore translates to $D \sim F L_C/\left(a \sqrt{1-\rho/\rho_*}\right)$, where $F$
    is the thickness of the profile $\tanh\left(x/2\right)$. We can have an estimate for $F$ from the spacing in between
    the two points where
    $\partial_x^3\tanh
    \left( x/2 \right) =0$. 
    This solves to $x=\ln\left(2\pm\sqrt{3}\right)\approx \pm 1.317$, meaning
    that $F\approx 2.63$. This corresponds to a the region where the field rises to $60\%$ of its asymptotic value. If we
    instead choose the region within which the field rises to $90\%$ of its asymptotic value, we are left with
    $F\approx 6$.
    $D$ is smallest in total vacuum, when $\rho/\rho_*\rightarrow 0$, and at the current
    time, when $a\rightarrow 1$. Since we are dealing with an inhomogeneous
    density distribution, the domain wall thickness will be environmentally dependent. 

    This means that within the steepest part, where the third derivative is positive, we will in the worst case have a resolution of 
    $D/\rm d x\approx 2.63\, L_C/\rm d x$ points along the wall profile. While,
    for the region where the field rises to $90\%$ of its asymptotic value, we will have
    $D/\rm d x\approx 6\, L_C/\rm d x$ points. In the majority of cases, and for the
    majority of the time of the simulation, the ambient background density will be larger so that
    $L_C/\sqrt{1-\left(\rho/\rho_*\right)^3}> L_C$, and $a<1$, in which case we will resolve more points. In our simulations, we have kept
    $L_C/\rm d x>1$. Simulations \rom{1} and \rom{3} are best and have $L_C/\rm d x\approx 2.6$. For these
    simulations we will in the worst case have $15$ points across the region where the wall rises
    to up to $90\%$ of its asymptotic value. For simulation \rom{1}, we also ran a lower resolution
    simulation, with $L_C/\rm d x = 1.024$, matching the resolution of simulations \rom{4} and \rom{5}, and found perfect agreement
    with the higher resolution run within the rising part of the power spectrum, see again figure \ref{fig:convergence}.

    The spatial and temporal resolution of the dynamical and nonlinear field through the phase transition
    and subsequent rearrangements is complicated to establish and is handled to much more detail in the
    previous work of \cite{christiansen_asimulation_2024}, whose rules we applied here, and to which we direct the interested reader for more details.

\end{document}